\documentclass[9pt,twocolumn,twoside]{opticajnl}
\journal{opticajournal} % use for journal or Optica Open submissions

\setboolean{shortarticle}{false}
\usepackage{lineno}

\usepackage{comment}
\usepackage{soul}
\usepackage{float}
\usepackage{hyperref}

\newcommand{\highlight}[1]{{\color{red} #1}}

\title{Full Kummer Beams: Skyrmionic textures in bichromatic light}

\author[1]{Jefferson Delgado-Quesada}
\author[1,2,3]{Miguel A. Alonso}
\author[1,4,*]{Eileen Otte}

\affil[1]{The Institute of Optics, University of Rochester, Rochester, NY 14627, USA.}
\affil[2]{Aix Marseille Univ, CNRS, Centrale Med, Institut Fresnel, UMR 7249, 13397, Marseille, Cedex 20, France.}
\affil[3]{Laboratory for Laser Energetics, University of Rochester, Rochester, NY 14627, USA.}
\affil[4]{Center for Coherence and Quantum Optics, University of Rochester, Rochester, NY 14627, USA.}

\affil[*]{eotte@ur.rochester.edu}

\begin{abstract} 
In this work, we introduce full Kummer beams, the bichromatic analog of full Poincaré beams. To this end, we define three parameters that characterize bichromatic states of polarization and span the Kummer shape, the bichromatic counterpart of the Poincaré sphere. We find that, under certain conditions, full Kummer beams can exhibit skyrmionic textures with a Skyrme number that depends not only on the topological charges but also on the frequencies. These beams can also exhibit emergent behavior upon propagation, including robust, higher-order-like structures and a propagation-invariant Skyrme number over a wide range of beam-waist selections.
\end{abstract}

\doi{}

\setboolean{displaycopyright}{false} % Do not include copyright or licensing information in submission.

\begin{document}

\maketitle

\section{Introduction}
Polarization of paraxial light is conveniently described by the Stokes parameters, which span the Poincaré unit sphere when normalized. For fully polarized light, a particular state of polarization is represented as a point on this surface, so a beam with spatially varying polarization can be visualized as a region of the sphere. In fact, full Poincaré beams have a cross section where each state of polarization appears at least once, thus covering the entire Poincaré sphere\,\cite{Beckley2010FullBeams}. Such beams (as well as the devices used to generate them) have proven beneficial in various fields of research, ranging from applications in imaging polarimetry\,\cite{Curcio2020BirefringentOrientation,Ramkhalawon2013ImagingField,Sivankutty2016Single-shotFiber} and material machining\,\cite{Omatsu2026NanofabricationSkyrmions}, to fundamental research in singular and topological optics\,\cite{Bauer2015ObservationStrips}. In recent years, full Poincar\'{e} beams have attracted particular attention as a realization of ``particle-like'', skyrmionic textures in light\,\cite{Gao2020ParaxialBeams,Cisowski2023BuildingMaps,Sugic2021Particle-likeLight,Ornelas2024Non-localLight,Marco2024PeriodicProjections,Marco2024Propagation-InvariantLattices,Marco20254DLight,Shen2025Free-spaceTutorial,Cheng2026NavigatingTutorial}. %  \JDQ{(move this to next paragraph)}. The Stokes vector, that is, the state of polarization, of these beams may exhibit a skyrmionic texture characterized by a topological invariant\,\cite{Gao2020ParaxialBeams}.

What we now know as a skyrmion, originally introduced by Skyrme\,\cite{Skyrme1961ATheory,Skyrme1962ABaryons}, is defined as a vector field with a topologically robust configuration. Its spatial distribution, referred to as a texture, covers all possible vector states within a region, thus spanning a compact parameter space. This skyrmionic texture is traditionally described by a mapping from 3D space to a 3-sphere and is characterized by a topological invariant. Although the concepts of skyrmion and skyrmionic texture go back to the nonlinear sigma model proposed to study particle physics phenomena, equivalent topological constructs have been realized in physical systems such as chiral magnetic materials, where textures manifest in the local orientation of the magnetic spin\,\cite{Nagaosa2013TopologicalSkyrmions,Gobel2021BeyondQuasiparticles,Tokura2020MagneticMaterials}. A more recent platform is optics, where skyrmionic textures have been shown to arise in vector properties of the electric field, such as the direction of the instantaneous field, the major axis of the polarization ellipse, or the spin vector\,\cite{Tsesses2018OpticalFields,Du2019Deep-subwavelengthMomentum,Gutierrez-Cuevas2021OpticalSpace,Koni2025Dual-WavelengthDefects}; besides being of fundamental interest, these structures are being explored for communication due to their potential robustness\,\cite{Peters2026TopologicalLink}. Note that, when the vector quantity is paraxial polarization, one does not observe a 3D skyrmion as defined in the traditional sense (projection from 3D space to a 3-sphere) but a 2D analog, namely, a ``baby skyrmion''.  In this case, the underlying mapping is between 2D space and a 2-sphere: the transverse plane of the electric field and the Poincaré sphere\,\cite{Ye2024Theoryskyrmions,Marco2024Propagation-InvariantLattices,Marco2024PeriodicProjections}.% \JDQ{(would it be more accurate to talk about mappings between two 2-spheres?)}.

%A currently unexplored order parameter space is that spanned by the quantities describing the polarization of a paraxial electric field with two commensurate frequencies, i.e., a bichromatic field. Such parameters span a surface known as the Kummer shape, which is a bichromatic analog to the Poincaré sphere\,\cite{Alonso2025GeometricCavities}. Previous studies on bichromatic light have focused on the singularities in superpositions of two commensurate monochromatic fields, each with spatially varying polarization, e.g., a full Poincar\'{e} beam undergoing second harmonic generation\,\cite{Kessler2003LissajousSingularities,Freund2003BichromaticFields,Freund2003PolarizationFields,Freund2003PolychromaticSingularities,Miao2022DesignBeams,Miao2023LissajousExperiment}. 

Skyrmionic textures constructed in the polarization of light are typically realized with monochromatic beams spanning all possible polarization ellipses. However, there are more complex and richer polarization orbits that can be generated with bichromatic light---these polarization states provide access to a variety of parameter spaces that have not been explored in the context of skyrmionic structures and offer a new playground for the generation of textures. Previous studies on bichromatic light have focused on the singularities in superpositions of two commensurate monochromatic fields, each with spatially varying polarization, e.g., a full Poincar\'{e} beam undergoing second harmonic generation\,\cite{Kessler2003LissajousSingularitiesb,Freund2003BichromaticFields,Freund2003PolarizationFields,Freund2003PolychromaticSingularities,Miao2022DesignBeams,Miao2023LissajousExperiment}. In the special case where each beam has spatially uniform polarization, the quantities describing bichromatic polarization span a surface known as the Kummer shape, which is a bichromatic analog of the Poincaré sphere\,\cite{Alonso2025GeometricCavities}. Exploring skyrmionic textures in this space offers control of new quantities such as the relative phase between two commensurate fields and their frequency ratio, as well as the emergence of new singularities, different topological characteristics, and distinct behavior under propagation. Motivated by this, we study skyrmionic textures in bichromatic light, specifically an electric field that consists of a superposition of two orthogonal components, each with spatially uniform polarization and different commensurate frequencies.

\begin{figure}[h]
    \centering
    \includegraphics[width=\columnwidth]{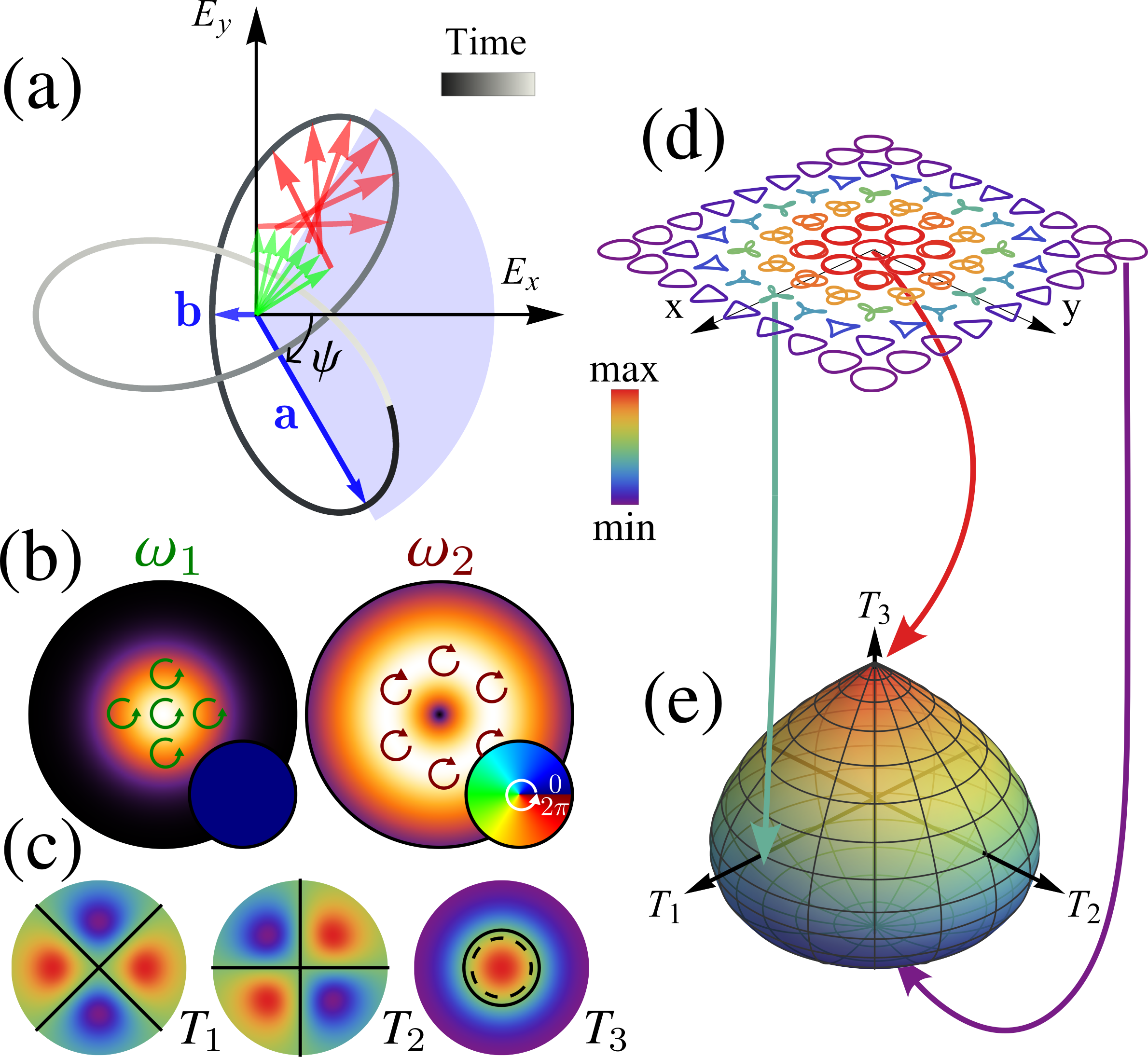}
    \caption{Graphical abstract of this work. (a) Time evolution of the orbit traced by the tip of the total electric field, with individual monochromatic components shown in green and red. Transparency corresponds to time evolution. The semi-major axis $\mathbf a$ (large blue arrow) defines the orientation $\psi$ of the orbit, and the sign of its cross product with the semi-minor axis $\mathbf b$ (small blue arrow) determines the handedness. The range of orientations lies within an arc of angle $2\pi/(m_1+m_2)$, shown in blue. (b) Pair of Laguerre-Gaussian beams superposed to generate bichromatic field. Spatial amplitude and phase profile of individual beams, consisting of fundamental Gaussian (left) and higher-order Laguerre-Gaussian (right) beams. (c) Bichromatic Stokes parameters with their zero-lines shown in black. Dashed line corresponds to the equator of the Kummer shape. (d) Orbits in transverse plane projected onto (e) Kummer shape. Color scheme of both (d) and (e) corresponds to the value of $T_3$, in rightmost (c).}% \JDQ{change green and purple arrows so they don not overlap and are curvy}}
    \label{fig:intro}
\end{figure}

A graphical overview of our work is shown in Fig.~\ref{fig:intro}. First, in Sec.~\ref{sec:stokes_and_kummer}, we introduce the fundamentals for describing bichromatic polarization states and the relationship between bichromatic Stokes parameters and geometric properties of orbits in the case of bicircular states, as depicted in Fig.~\ref{fig:intro}a. Subsequently, in Sec.~\ref{sec:kummer_beams}, we generate full Kummer beams by superposing two Laguerre-Gaussian beams with commensurate frequencies (Fig.~\ref{fig:intro}b), thereby spanning the Kummer shape and covering all possible values of the bichromatic Stokes parameters (Fig.~\ref{fig:intro}c). Finally, in Sec.~\ref{sec:general_properties}, we investigate the requirements for a nontrivial Skyrme number characterizing the mapping between the transverse plane of the bichromatic field and the Kummer shape, as shown in Figs.~\ref{fig:intro}d, e.

\section{Bichromatic Polarization States}
\label{sec:stokes_and_kummer}
%$\mathbf E =\Re\left[\mathbf{\hat e}_1 E_1 + \mathbf{\hat e}_2 E_2\right]=\Re\left[\mathbf{\hat e}_1E_{01}e^{-i\omega_1t}+\mathbf{\hat e}_2E_{02}e^{-i\omega_2t}\right]$
Consider the superposition of two mutually coherent scalar monochromatic electric fields, whose complex (analytic signal) representations are
\begin{equation}
    E_1=E_{1}^0\exp{(-\mathrm{i} \omega_1t)} \qquad \text{and} \qquad E_2=E_{2}^0\exp{(-\mathrm i\omega_2t)} 
    \label{eq:scalarorthogonalbeams}
\end{equation}
with different frequencies $\omega_{1,2}$, complex amplitudes $E_{1,2}^0=C_{1,2}\exp(\mathrm i \gamma_{1,2})$, and orthonormal states of polarization $\mathbf{\hat e}_{1,2}$. We assume that $\omega_1$ and $\omega_2$ are commensurate, i.e., $\omega_2/\omega_1=m_2/m_1$ where $m_1$ and $m_2$ are coprime positive numbers, typically of modest magnitude. Unlike the polarization ellipse for monochromatic light, the electric field [real part of Eq.~(\ref{eq:scalarorthogonalbeams})] in such a bichromatic field traces more complex orbits over time. In the context of bichromatic light, these orbits are commonly referred to as Lissajous shapes (which include ellipses as a special case for $\omega_1=\omega_2$), each shape corresponding to a different {\it bichromatic polarization} state of the total electric field\,\cite{Kessler2003LissajousSingularitiesb,Gbur2016SingularOptics}.% Notice that the bichromatic polarization state relates to the total electric field, but each monochromatic component $E_{1,2}$ possesses a defined monochromatic polarization, $\mathbf{\hat e}_1$ and $\mathbf{\hat e}_2$.

\begin{figure}[h]
    \centering
    \includegraphics[width=0.8\columnwidth]{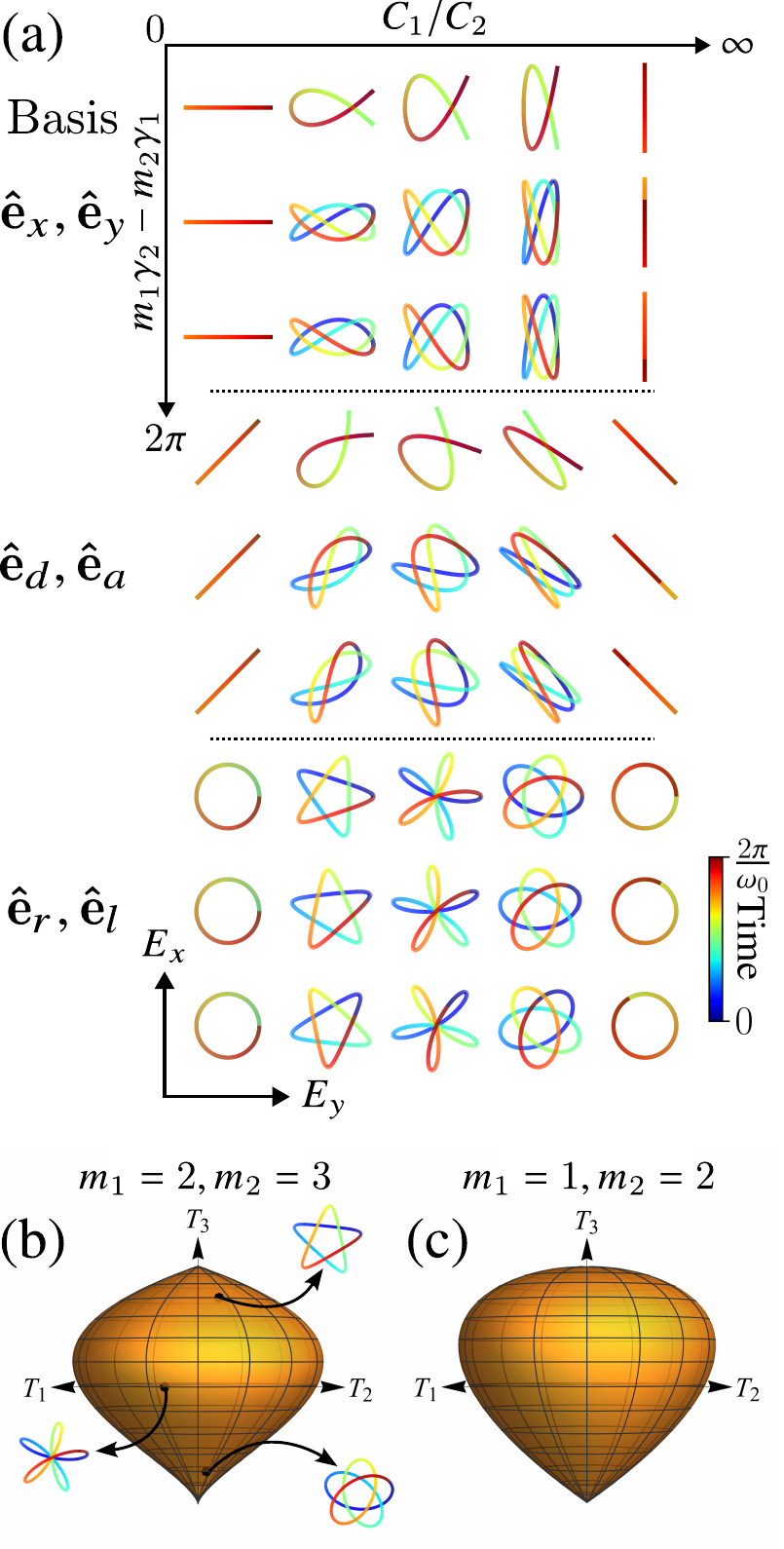}
    \caption{Orbits of bichromatic polarization states and Kummer shapes. (a) Exemplary orbits for $m_1=2,m_2=3$ in the three canonical bases: horizontal and vertical (top), diagonal and antidiagonal (middle), right and left circular (bottom). The color scheme of all orbits indicates time evolution during a period of the fundamental frequency defined as $\omega_0=\omega_1/m_1=\omega_2/m_2$. Relative amplitudes (phases) are changed from left to right (top to bottom). Kummer shapes for (b) $m_1=2,m_2=3$ and (c) $m_1=1,m_2=2$, where some orbits at the corresponding points are shown in (b).}
    \label{fig:example_orbits}
\end{figure}

Examples of such trajectories are shown in Fig.~\ref{fig:example_orbits}a for $m_1=2,m_2=3$ and for different choices of $\mathbf{\hat e}_{1,2}$, where the color scheme of all orbits indicates time evolution during a period of the fundamental frequency, defined as $\omega_0=\omega_1/m_1=\omega_2/m_2$. Fields $E_1$ and $E_2$ with different amplitudes $C_{1,2}$ (cf. Fig.~\ref{fig:example_orbits}a, left to right) or phases $\gamma_{1,2}$ (cf. Fig.~\ref{fig:example_orbits}a, top to bottom) may produce a different bichromatic polarization state, in the same manner that varying the relative amplitude and phase shifts between the electric field components of monochromatic light alters the polarization ellipse, hence, the monochromatic state of polarization.

\subsection{Bichromatic Stokes Parameters and Kummer Shape}
For monochromatic light, we find different states of polarization conveniently quantified through Stokes parameters. However, for bichromatic light, standard Stokes parameters do not adequately represent bichromatic polarization states: they only capture the phase difference $\gamma_2-\gamma_1$, and Lissajous shapes are invariant not only under phase shifts of $2\pi$ but also under the transformation $\gamma_{1,2}\rightarrow\gamma_{1,2}+2\pi/(m_{2,1})$, which is not reflected by the quantity $\gamma_2-\gamma_1$. Hence, we define {\it normalized bichromatic Stokes parameters}\,\cite{Alonso2025GeometricCavities,Holm2011GeometricEdition}
\begin{subequations}
\label{eq:bi_stokes}
\begin{align}
    \label{eq:Tb}
    T_1&=\frac{C_{m_1m_2}}{\tau_0^{(m_1+m_2)/2}}\Re[(E_1^*)^{m_2}E_2^{m_1}],\\%\highlight{\rightarrow \frac{2}{\tau_0}\frac{\Re[(E_1^*)^{m_2}E_2^{m_1}]}{|E_1|^{m_2-1}|E_2|^{m_1-1}}}, \\
    \label{eq:Tc}
    T_2&=\frac{C_{m_1m_2}}{\tau_0^{(m_1+m_2)/2}}\Im[(E_1^*)^{m_2}E_2^{m_1}], \\%\highlight{\rightarrow \frac{2}{\tau_0}\frac{\Im[(E_1^*)^{m_2}E_2^{m_1}]}{|E_1|^{m_2-1}|E_2|^{m_1-1}}}, 
    \label{eq:Ta}
    T_3&=\frac{1}{\tau_0}(|E_1|^2-|E_2|^2),
\end{align}
\end{subequations}
where $\tau_0=|E_1|^2+|E_2|^2$ is proportional to the time-averaged irradiance and $C_{m_1m_2}=\sqrt{(m_1+m_2)^{m_1+m_2}/(m_1^{m_1}m_2^{m_2})}$ is a normalization factor that ensures all three parameters are bounded between $-1$ and $1$, as known from monochromatic Stokes parameters. The non-bilinearity of Eqs.~(\ref{eq:Tb}) and (\ref{eq:Tc}) makes the product in square brackets a constant of motion by appropriately canceling out the time dependence of each monochromatic component of the complex electric field\,\cite{Alonso2025GeometricCavities}. These three quantities uniquely represent a given bichromatic polarization state through the vector $\mathbf T=(T_1,T_2,T_3)$. Note that the components of $\mathbf T$ do not cycle under transformations of the polarization basis the way monochromatic Stokes parameters do; hence, their validity is tied to the choice of $\mathbf{\hat e}_1$ and $\mathbf{\hat e}_2$.

For monochromatic light, Stokes parameters span the Poincar\'{e} sphere, given their relation $s_0^2=s_1^2+s_2^2+s_3^2$ (with $s_0=1$ for normalized parameters). In contrast, for bichromatic light, given the full coherence between the two underlying monochromatic components, one finds: %that \MAA{\it (A minor grammatical detail: better to remove the "that" because leaving it means the $=$ sign in the equation that follows has to act as a verb, and it is better not to do this)}
\begin{equation}
    \label{eq:kummer_shape}
    T_1^2+T_2^2 = C_{m_1m_2}^2\left(\frac{1-T_3}{2}\right)^{m_1} \left(\frac{1+T_3}{2}\right)^{m_2}.
\end{equation}
Therefore, only two of the three bichromatic Stokes parameters are actually independent, which becomes explicit with the following convenient parameterization: %\MAA{\it (In 4a, use the two-argument version of arctan, which is defined within $(-\pi,\pi]$? Also, notice that the limits of this periodic interval cannot be inclusive on both ends)}
\begin{subequations}
\label{eq:kummer_params}
\begin{gather}
    \label{eq:varphi}
    \varphi  = \mathrm{atan2}{\left(T_2,T_1\right)}\in[-\pi,\pi),\\% \in [0,2\pi].
    \label{eq:theta}
    %\theta &= \arccos{(T_3)}, \\%\in[0,\pi], \\
    T_3\in[-1,1].
\end{gather}
\end{subequations}
The surface defined by Eq.~(\ref{eq:kummer_shape}) is known as the Kummer shape\,\cite{Holm2011GeometricEdition}. This shape forms a bichromatic analog of the Poincar\'{e} sphere, where different points on the surface correspond to different bichromatic polarization states. Its specific form depends on $m_1$ and $m_2$, as shown in Fig.~\ref{fig:example_orbits}b, c. Some characteristics are common regardless of the value of $m_1$ and $m_2$: the surface has rotational symmetry with respect to the $T_3$ axis, and its widest cross section, which we refer to as the equator, is located at
\begin{equation}
    T_3=\epsilon\equiv\frac{m_2-m_1}{m_1+m_2}.
\end{equation} 
Moreover, the north and south poles, situated at $\mathbf T=(0,0,\pm1)$, are the only points on the Kummer shape whose orbits correspond to ellipses, which are associated with $\mathbf{\hat e}_{1,2}$, that is, the polarization states of the individual beams in Eq.~(\ref{eq:scalarorthogonalbeams}). The surface at the north (south) pole is not smooth when $m_1>1$ ($m_2>1)$, which means that one of the poles is smooth only when one of the frequencies is an integer multiple of the other, e.g., a fundamental frequency and one of its harmonics (such as the case shown in Fig.~\ref{fig:example_orbits}c). Note that Eqs.~(\ref{eq:bi_stokes}) and (\ref{eq:kummer_shape}) simplify to the normalized monochromatic Stokes parameters and the Poincar\'{e} sphere when $\omega_1=\omega_2$ (i.e., $m_1=m_2=1$).

The reason the Kummer shape is not a sphere comes from the non-bilinearity of Eqs.~(\ref{eq:Tb}) and (\ref{eq:Tc}), which manifests on the right hand side of Eq.~(\ref{eq:kummer_shape}). In order to instead generate a sphere, one could either modify $T_{1,2}$ or $T_3$, enabling the mapping of the Kummer shape onto a sphere. % The reason Eq.~(\ref{eq:kummer_shape}) is not a sphere comes from the non-bilinearity of Eqs.~(\ref{eq:Tb}) and (\ref{eq:Tc}). Non-bilinearity is only relevant for the phase of $T_1+\mathrm iT_2$, but not for its amplitude, so one could also define the quantities $T'_{1,2}= 2\tau_0^{(m_1+m_2)/2-1}T_{1,2}/(C_{m_1m_2}|E_1|^{m_2-1}|E_2|^{m_1-1})$ and $T_3'=T_3$, which satisfy $T_3'^2+T_1'^2+T_2'^2=1$, i.e., span a sphere. 
However, as we will point out in the following sections, the poles of the Kummer shape exhibit a special behavior that is explicitly highlighted by their pinched aspect; this behavior turns out to be connected to the singular properties of the field formed when a region of the Kummer shape containing one of these poles is mapped onto a cross section of the beam.%when the poles are mapped onto a cross section of the beam!!!!!!} \MAA{SAY INSTEAD: this behavior turns out to be connected to the singular properties of the field when a region of the Kummer shape containing one of these poles is mapped onto a cross section of the beam?}

%this makes the Kummer shape the right surface to adequately describe the properties of bichromatic states of polarization. \MAA{\it (Maybe we will have to come back to discuss this.)}

We know that a point on the Poincaré sphere relates to the physical and geometrical properties of the respective polarization ellipse. Specifically, the orientation of the major axis and the ellipticity are quantified by the azimuthal and polar angles, respectively. These relations are not immediately evident for bichromatic light, its polarization orbits, and the Kummer shape. In the following, we derive such relations. %In the following section, we concentrate on the case of circular polarization basis, also known as bicircular light, to define such relations.

\subsection{Orientation and Handedness of Bicircular States}%Circular basis}
%the orientation is no longer defined because it is calculated as one of the eigenvectors of the real coherency matrix. This matrix is simply diagonal when $\mathbf{\hat e}_1$ and $\mathbf{\hat e}_2$ are uniform since the time average of $E_xE_y$ is identically zero and, yielding only two possible orientations: horizontal and vertical. Regarding handedness, Freund's definition does not vanish, but their definition is only valid for $m_1=1,m_2=2$; a generalization is therefore needed for arbitrary $m_1,m_2$.
\label{sec:orientation_and_handedness}
%\MAA{\it (Open the section explaining the circular basis. The title is not enough to introduce the fact that we are using this basis.) \JDQ{JDQ: since major/minor axes are more general, I started with that, then bicircular states.}}\\%% \MAA{\it (We might have discussed this already... is this the rl convention that Eileen (or Greg or Freund) uses? I use the opposite one in my papers :-) !) \JDQ{JDQ: this is the convention I like and used in calculations hehe, I could change it if needed}} 
For monochromatic light, the major and minor axes of the polarization ellipse represent key tools in defining the orientation and handedness of a state of polarization. Obviously, those axes do not readily translate to the significantly more complex orbits of bichromatic polarization. To find an analogous tool, we study bicircular states, which consist of a superposition of two components with circular but orthogonal polarizations, that is, $\mathbf{\hat e}_1=\mathbf{\hat e}_\mathrm L=(\mathbf{\hat e}_x+\mathrm i\mathbf{\hat e}_y)/\sqrt{2}$ and $\mathbf{\hat e}_2=\mathbf{\hat e}_\mathrm R=(\mathbf{\hat e}_x-\mathrm i\mathbf{\hat e}_y)/\sqrt{2}$. We will use vectors $\mathbf a$ (semi-major axis) and $\mathbf b$ (semi-minor axis), being the consecutive maximum and minimum electric fields, respectively, of an orbit. The extrema are defined by the magnitude of the electric field, which simplifies to
\begin{equation}
    %\mathbf E = \frac{1}{\sqrt{2}}\left[A\cos(\omega_1t-\alpha) + B\cos(\omega_2t-\beta)\right]\mathbf{\hat x} \nonumber\\
    %+\frac{1}{\sqrt{2}}\left[A\sin(\omega_1t-\alpha) - B\sin(\omega_2t-\beta)\right]\mathbf{\hat y}\\
    %\Rightarrow 
    |\mathbf E|^2 = \tfrac{1}{2}\left(C_1^2+C_2^2\right)+C_1C_2\cos{[(\omega_1+\omega_2)t-\gamma_1-\gamma_2]}.
\end{equation}
% where $C_1$ and $\gamma_1$ ($C_2$ and $\gamma_2$) are the amplitude and phase of $E_{1}^0$ ($E_{2}^0$).
Following the cosine term, the magnitude reaches a maximum when $(\omega_1+\omega_2)t-\gamma_1-\gamma_2=2j\pi$, followed by a minimum when $(\omega_1+\omega_2)t-\gamma_1-\gamma_2=(2j+1)\pi$ with $j$ being an integer. Let the orientation of the bichromatic polarization state be defined by $\psi$, the angle formed by the semi-major axis $\mathbf a $ with respect to the $x$ axis; when $j=0$, it is: %Let $\mathbf a$ and $\mathbf b$ be consecutive maximum and minimum electric fields, respectively, corresponding to $j=0$. %These two vectors now act as generalizations of the semi-major and semi-minor axes of the polarization ellipse for monochromatic light. The generalized major axis $\mathbf{a}$ can be used to define the orientation of bichromatic states as the angle between $\mathbf{a}$ and the $x$ axis, given by
\begin{equation}
    \psi \equiv \arctan\left(\frac{a_y}{a_x}\right)= \frac{m_1\gamma_2-m_2\gamma_1}{m_1+m_2},%=\frac{2\psi^u}{m_1+m_2},=\frac{\omega_1\beta-\omega_2\alpha}{\omega_1+\omega_2}
\end{equation}
%\highlight{
%\begin{equation}
%    \frac{a_y}{a_x}=\frac{\displaystyle \sin\left(\frac{m_1\beta-m_2\alpha+2nm_1\pi}{m_1+m_2}\right)}{\displaystyle \cos\left(\frac{m_1\beta-m_2\alpha-2nm_1\pi}{m_1+m_2}\right)}
%\end{equation}
%physically, one particular axis cannot be tracked...
%}
resulting in
\begin{equation}
    \label{eq:orientation_circular}
    \tan[(m_1+m_2)\psi]=\frac{T_2}{T_1}\Leftrightarrow \varphi=(m_1+m_2)\psi.%=\frac{S_2^u}{S_1^u},
\end{equation}
% \MAA{\it (Note here that $\varphi$ is proportional to $\psi$, generalizing the paraxial result.)} 
Crucially, this means that the orientation of bichromatic, bicircular polarization is directly related to the ratio of the second and first bichromatic Stokes parameters---similar to how the orientation of the polarization ellipse for monochromatic light is given by the ratio of the second and first standard Stokes parameters. Equation~(\ref{eq:orientation_circular}) also shows that the orbits are invariant under rotations of $2\pi/(m_1+m_2)$ radians, thus making orientation not uniquely defined and being the reason a particular index $j$ must be chosen. This does not represent an issue, however, since an angle in the range $\psi\in\left[ -\pi/(m_1+m_2),\pi/(m_1+m_2)\right]$ always matches the location of one of the $m_1+m_2$ maxima of the orbit. Note that the same situation occurs in the monochromatic case, where orientation of the polarization ellipse is only defined between $-\pi/2$ and $\pi/2$.
%where one of the maxima will always be found.

%\EO{... (write something about the m12 dependency and the rotational symmetry of states)}.

%The derivation of Eqs. (\ref{eq:orientation_circular}) and (\ref{eq:h_geo_circular}) was carried out considering a specific value of $j$. Although the latter holds for arbitrary $j$, the former requires $j=0$. This stems from  \MAA{\it (Move this equation to the previous paragraph?)}\EO{yes, I agree, see my comments}

%\MAA{\it (Maybe bring everything related to the relations between the angles, the magnitudes $a$ and $b$, and the first two Stokes parameters to the previous paragraph, and open a new paragraph discussing that $T_3$ (and therefore $\theta$ are related to areas and spin in the monochromatic case, so we will explore if this generalizes to the bichromatic one. Somewhere in the paragraph you can maybe mention also that $T_3=2|\mathbf{a}||\mathbf{b}|/(|\mathbf{a}|^2+|\mathbf{b}|^2)$ like in monochromatic case but geometrical interpretation trickier?) \JDQ{JDQ: I want to confirm how exactly you want to move things around before doing it.}} 
To define the handedness of a bichromatic state, we can use the signed area of the parallelogram formed by the semi-major and semi-minor axes, 
%On the other hand, the signed area of the parallelogram formed by the semi-major and semi-minor axes is 
%
\begin{align}
    \label{eq:Aparallel}
    A_\mathrm{parallel}&=\mathbf z\cdot(\mathbf a\times \mathbf b)=\frac{1}{2}\sin{\left(\frac{m_1}{m_1+m_2}\pi\right)}(C_1^2-C_2^2) \nonumber \\
    &=\frac{\tau_0T_3}2\cos\left(\epsilon\frac{\pi}2\right).
\end{align}
This signed area characterizes the sense (clockwise or counterclockwise) in which the electric field rotates around the origin. Hence, it can be used to geometrically define handedness as
\begin{equation}
    \label{eq:h_geo_circular}
    h_\mathrm{geo}=\mathrm{sign}~A_\mathrm{parallel}=\mathrm{sign}~T_3.%=\mathrm{sign}~S_3^u
\end{equation}
%\MAA{{\it Maybe the following sentence is no longer needed?}} In contrast, the sign of the cross product in Eq.~(\ref{eq:h_geo_circular}) will remain the same for any pair of consecutive maxima and minima.
%\MAA{\it (Would it be better instead then to do the derivation general? One would have to just let $m$ be any integer but force $n=m$ for Emin to be the minimum that follows immediately after Emax. They could be defined as consecutive from the beginning, so we do not need $n$?)} 
An alternative quantity that could be used to define handedness is the time-averaged signed area swept by the electric field; this approach is analogously applicable for monochromatic light\,\cite{Berry2004IndexPolarization,Freund2003BichromaticFields}. The swept area can be interpreted as the angular velocity of the tip of the electric field, therefore determining its sense of rotation, and is given by%\EO{(cite Freunds paper on lissajous; Berrys definition of p and q for nonparaxial light, he also uses this for handedness)} \MAA{(\it connect with SAM? \JDQ{JDQ: what if one uses the area swept by the vector potential?})}. 
\begin{align}
    \label{eq:Aswept}
    A_\mathrm{swept} &= \lim_{t'\rightarrow\infty}\frac{1}{t'}\int_0^{t'} \mathrm d t\left[\mathbf{\hat z}\cdot\left(\mathbf E\times\frac{\partial \mathbf E}{\partial t}\right)\right] = \tfrac{1}{2}(\omega_1C_1^2-\omega_2C_2^2) \nonumber \\
    &= \tfrac{1}{4}\tau_0(\omega_1+\omega_2)(T_3-\epsilon),
\end{align}
leading to the handedness
\begin{equation}
    \label{eq:h_time_circular}
    h_\mathrm{ave} = \mathrm{sign}~A_\mathrm{swept} =\mathrm{sign}(T_3-\epsilon).
\end{equation}
For both approaches, Eqs.~(\ref{eq:h_geo_circular}) and (\ref{eq:h_time_circular}), the handedness is directly related to the third bichromatic Stokes parameter $T_3$; this again reveals a similarity to the monochromatic case where handedness is related to Stokes parameter $s_3$. Note that vanishing handedness has a different meaning for each definition and occurs at distinct $T_3$ locations: $h_\mathrm{geo}$ is zero when the semi-minor axis $\mathbf b$ is undefined, which happens at $T_3=0$; $h_\mathrm{ave}$ is zero when the areas swept by the electric field clockwise and counterclockwise are equal on average, which corresponds to the equator of the Kummer shape (i.e., $T_3=\epsilon$). In addition, both areas, Eqs.~(\ref{eq:Aparallel}) and (\ref{eq:Aswept}), can be related to spin, as shown in the Supplementary Information.% That is, there is no unique measure of spin (and hence handedness) for this type of bichromatic field, as there are at least two reasonable options [given in Eqs.~(\ref{eq:Aparallel}) and (\ref{eq:Aswept})] that depend not only on $T_3$ but also on $\epsilon$.

%Interestingly, since bichromatic Stokes parameters do not have a unique definition, they could be modified so that the sign of Eq. (\ref{eq:h_time_circular}) equals the sign of $T_3$.

%\JDQ{(In the monochromatic case, $\varphi$ and $\theta$ correspond, respectively, to $2\psi$ and $\pi/2-2\chi$. A natural question is: is there an equivalent here? For the first case, it would be $(m_1+m_2)\psi$. I have not found anything for the second case.)} \MAA{\it See comments above.}

\section{Kummer Beams and Full Kummer Beams}
\label{sec:kummer_beams}
By using the tools presented above, we can now introduce Kummer and full Kummer beams as the bichromatic analogs of monochromatic Poincaré and full Poincar\'{e} beams, and study their skyrmionic properties. (Not to be confused with the Fraunhofer diffraction patterns produced by a binary phase vortex hologram, which some authors have also termed Kummer beams\,\cite{Pasko2004AnalysisCharge}.) Based on the established approach of creating monochromatic structured vectorial beams, we form spatially structured bichromatic fields by the superposition of two spatial modes with orthogonal polarization. Hence, in Eq.\,(\ref{eq:scalarorthogonalbeams}), the complex amplitudes $E_{1,2}^0$ become spatially structured, namely, Laguerre-Gaussian (LG$_{n \ell}$) beams \cite{Kogelnik1966LaserResonators}. This exact analytic solution of the paraxial, scalar wave equation in cylindrical coordinates $(\rho,\phi,z)$ represents the family of modes that was first used to introduce full Poincaré beams\,\cite{Beckley2010FullBeams}. Here, $\phi$ is the angle within the transverse plane, not to be confused with the angle $\varphi$ over the Kummer shape. Analogously, we %We will explore how the bichromatic Stokes parameters characterize the bichromatic polarization distribution of Laguerre-Gaussian beams, the family of light fields that was first used to introduce full Poincaré beamsAlonso2011FullBeams}. %\st{a specific family of light fields.} \MAA{\it (Make connection here or in introduction to the FPB, so the choice of beams below is justified.)} 
consider a fundamental Gaussian mode $\text{LG}_{00}$ and a higher-order mode $\text{LG}_{0\ell}$ with radial index $n=0$ and topological charge $\ell$\,\cite{Gbur2016SingularOptics,Siegman1986Lasers},% \MAA{\it (add other references for LG, such as Siegman?)}, 
\begin{align}
    E_{1}^0(\rho,z) &=\text{LG}_{00}= \frac{\sqrt{I_1}}{\xi_1}\exp{\left(-\frac{\rho^2}{w_1^2\xi_1^2}\right)}\nonumber\\
    &\times\exp{\left[\mathrm i\left(k_1z+\frac{k_1\rho^2}{2R_1}-\Phi_{\mathrm G1}\right)\right]},\\
    E_{2}^0(\rho,\phi,z) &= \text{LG}_{0\ell} =\frac{\sqrt{I_2}}{\xi_2
    }  \left(\frac{\sqrt{2}\rho}{w_2\xi_2}\right)^{|\ell|}\exp{\left(-\frac{\rho^2}{w_2^2\xi_2^2}\right)} \nonumber\\
    &\times\exp{\left\{\mathrm i\left[k_2z+\frac{k_2\rho^2}{2R_2}+\ell\phi-(|\ell|+1)\Phi_{\mathrm G2}\right]\right\}}.
\end{align}
For each of the two beams $j=\{1,\,2\}$, $I_j$ is a normalization constant, the parameter $\xi_j(z)=\sqrt{1+z^2/z_{{\mathrm R}j}^2}$ accounts for the natural hyperbolic expansion of the beams under propagation in the $z$ direction, $R_j(z)=z+z_{{\mathrm R}j}^2/z$ is the wavefront's radius of curvature, $\Phi_{{\mathrm G}j}(z)=\arctan(z/z_{\mathrm Rj})$ is the Gouy phase for the fundamental Gaussian, $k_j=\omega_j/c$ is the wavenumber (with $c$ being the speed of light), $z_{{\mathrm R}j}=k_j w_j^2/2$ is the corresponding Rayleigh range, and $w_j$ is the beam waist.

%\JDQ{[I changed the order below. Better now? (Miguel suggested introducing Skyrme density first then Skyrme number)]}% \JDQ{[Mathematically, this is referred to as a surjection or epimorphism (apparently)]}. %The specific way in which the Kummer shape is covered matters, and it must be characterized; 
The spatially varying amplitude and phase profile of these beams produces a bichromatic structured beam whose bichromatic Stokes parameters cover a portion of the Kummer shape. Thus, we refer to them as \textit{Kummer beams}. Within this family, a \textit{full Kummer beam} maps all states of polarization onto the Kummer shape at least once. We characterize the coverage using the Skyrme density $\Sigma$, which is a Jacobian that characterizes the ratio of the infinitesimal coverage of the compact parameter space (the Kummer shape) and that of the transverse plane\,\cite{Shen2025Free-spaceTutorial}. There are several options for quantifying the coverage of the Kummer shape and, hence, the Skyrme density, such as the area or the solid angle subtended by ${\bf T}$; a third option is based on the area over the projection of the Kummer shape onto the co-axial cylinder of unit radius that contains it\,\cite{Alonso2025GeometricCavities}. While these three options are equivalent for the monochromatic case (where the Kummer shape reduces to the Poincar\'e sphere), they differ in the bichromatic case. Here we choose the third option given that it leads to simpler expressions,
\begin{figure*}[t]
    \centering
    \includegraphics[width=\linewidth]{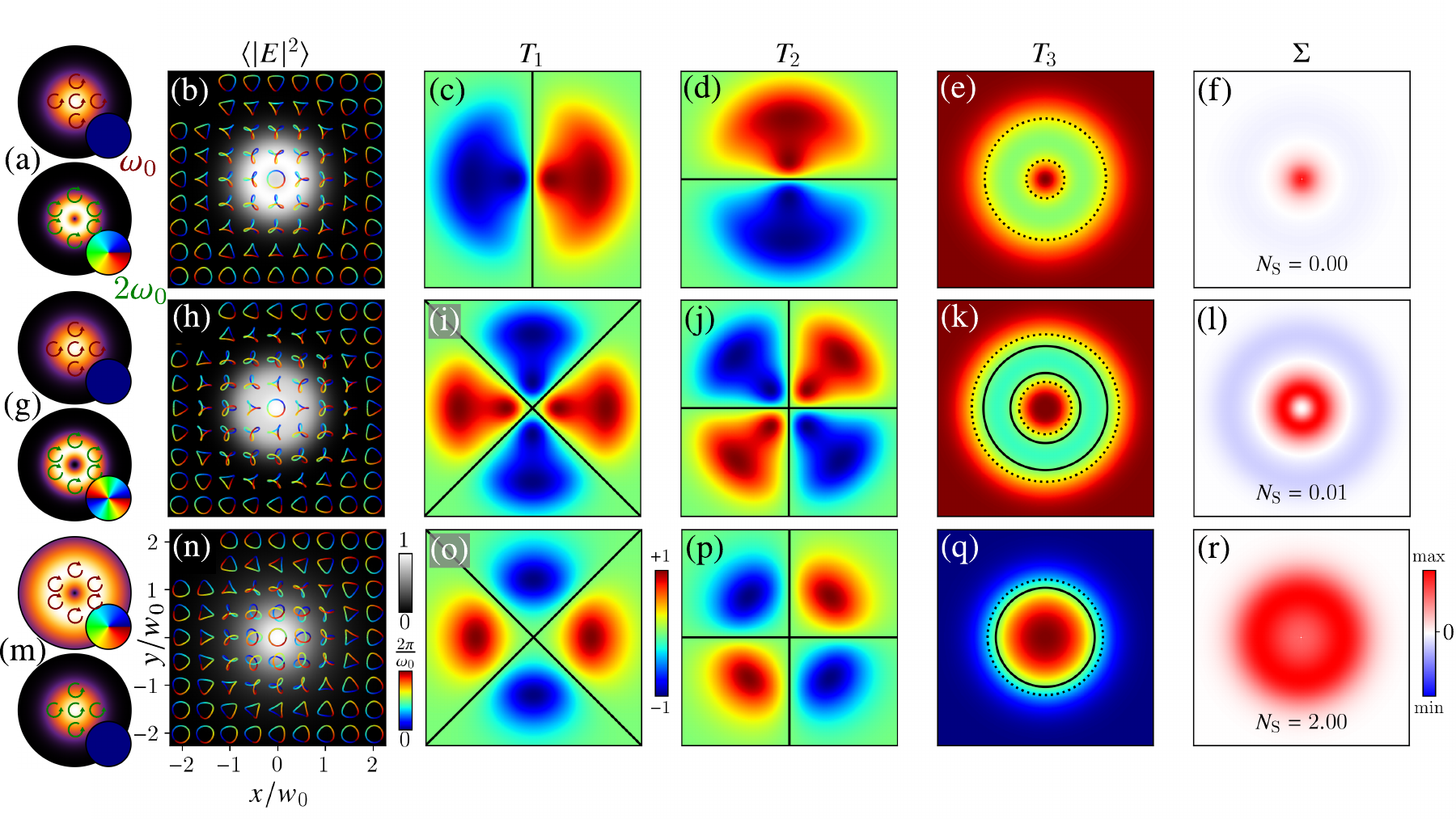}
    \caption{Examples of bichromatic Kummer beams. (col. 1) Spatial irradiance and phase profiles of superimposed individual beams, (col. 2) normalized time-averaged irradiance and bichromatic polarization distribution of the total electric field, (cols. 3-5) bichromatic Stokes parameters $T_{1,2,3}$, and (col. 6) Skyrme density, for (a-f) $m_1=1$, $m_2=2$, $\ell=1$, (g-l) $m_1=1$, $m_2=2$, $\ell=2$, and (m-r) $m_1=2$, $m_2=1$, $\ell=1$. Black solid lines in cols. 3-5 correspond to the zero-lines per parameter, and the black dashed line in col. 5 represents the equator of the Kummer ($T_3=\epsilon$). The Skyrme number is shown in col. 6. Rayleigh ranges are equal for all cases, and the beams have not propagated, i.e., $z=0$.}
    \label{fig:irradiance_and_stokes}
\end{figure*}
\begin{equation}
    \label{eq:skyrme_density}
    \Sigma = \mathbf n\cdot\left(\frac{\partial\mathbf n}{\partial x}\times\frac{\partial\mathbf n}{\partial y}\right) = \frac{\partial \varphi}{\partial x}\frac{\partial T_3}{\partial y} - \frac{\partial \varphi}{\partial y}\frac{\partial T_3}{\partial x},
\end{equation}
where $\mathbf n=(\cos\varphi,\sin\varphi,T_3)^\text T$ is a vector spanning the co-axial cylinder of unit radius that contains the Kummer shape. Notice that the Skyrme density $\Sigma$ is the Jacobian of the mapping $(x,y)\to(\varphi,T_3)$. From this, we calculate the Skyrme number $N_\text{S}$, which quantifies how many times the vector $\mathbf T$ wraps around the Kummer shape within a region $\mathcal R$ of the transverse plane. This number is given by\,\cite{Han2017skyrmionsMatter}
\begin{equation}
    \label{eq:skyrme_number}
    N_\mathrm S = \frac{1}{4\pi} \iint_{\mathcal R} \mathrm dx\mathrm dy~\Sigma.
\end{equation}
We refer to full Kummer beams as beams with \textit{skyrmionic textures} if they yield a nonzero integer Skyrme number within any region $\mathcal{R}$. % \highlight{(really any region???)}.
%A particular type of full Kummer beam is one that yields a nonzero integer Skyrme number within any region $\mathcal R$. We refer to such light fields as beams with skyrmionic textures.

%Mathematically, a Full Kummer beam can be defined as a light field . However, we may refer to it as a skyrmionic Kummer beam only if the region $\mathcal R$ is $\mathbb R^2$, i.e., the entire transverse plane. \JDQ{(change this... )}

%one region (finite or not) of the transverse plane mapping the entire Kummer shape. In addition, a specific type of full Kummer beam is a skyrmionic Kummer beam, whose entire transverse plane maps the whole Kummer shape in a specific way \JDQ{(qualitative description of this ``specific way''?)}.

%\MAA{\it (I inserted text in the discussion that follows regarding the definition of the Skyrme density, in terms of Kummer shape area, solid angle, or cylinder area. However I realized that when using area (which we don't) the normalization in Eq. (15) would be wrong since the Kummer shape's area is not $4\pi$, as opposed to the solid angle and the cylinder. Since we end up opting for the cylinder, I wonder if it is easier to discuss and define first the Skyrme density and wait to define the Skyrme number afterwards.)}

In Fig.~\ref{fig:irradiance_and_stokes}, we present examples of Kummer beams, illustrating all the quantities defined up to this point. We consider the fundamental Gaussian beam to be of left circular polarization, while the other beam has right circular polarization. Moreover, we have the combination of a fundamental frequency $\omega_0$ and its second harmonic $2\omega_0$. Between rows in Fig.~\ref{fig:irradiance_and_stokes}, values of $m_1$, $m_2$, and $\ell$ are varied, but the Rayleigh ranges of both beams are equal for all. The first column shows the spatial irradiance (left) and phase profiles (right inset) of the two individual, monochromatic beams, as well as their polarization (arrows on irradiance). The top (bottom) beam always corresponds to the frequency $\omega_0$ ($2\omega_0$). In the second column, we illustrate the bichromatic structured field resulting from the superposition of the two monochromatic beams. We show the normalized time-averaged irradiance (background) with bichromatic polarization distribution (orbits) of the total electric field.

The respective bichromatic Stokes parameters $T_{1,2,3}$ are shown in columns three to five, with black lines marking their zero-lines per parameter and the dashed line in $T_3$ highlighting the Kummer shape's equator, $T_3=\epsilon$. As discussed in Sec~\ref{sec:orientation_and_handedness}, $T_1$ and $T_2$, depicted in the second and third columns, together determine orientation. In general, orientation is only defined within the angles $-\pi/(m_1+m_2)$ and $\pi/(m_1+m_2)$, which correspond to $-60^\circ$ and $60^\circ$ for $m_1=1$ and $m_2=2$ (or vice versa). When $T_1=+1$ and $T_2=0$, the semi-major axis $\mathbf a$ is aligned with the $x$ axis, i.e., the orientation $\psi$ is zero. If $T_1=-1$ and $T_2=0$, the orientation is $\psi=\pm 60^\circ$; these two orientations are the same due to the three-fold symmetry of the orbits. The values $T_1=0$ and $T_2=\pm1$ represent the orientations $\psi=\pm30^\circ$. These four values are the bichromatic equivalents of horizontal, vertical, diagonal, and antidiagonal orientations. 

Similarly to the monochromatic case, intersections of the zero-lines of $T_1$ and $T_2$ correspond to points where orientation is undefined, which only occurs for circular orbits located at the poles of the Kummer shape. In fact, orbits with undefined orientation are $m$-times wrapped circles, where $m=m_1$ ($=m_2$) when they correspond to the north (south) pole of the Kummer shape; this visually appears in $T_1$ and $T_2$ as zero-lines intersecting multiple times. For instance, in Figs.~\ref{fig:irradiance_and_stokes}b, h (where $m_1=1$ and $m_2=2$), the origin corresponds to the north pole, and circular orbits there rotate only once. Note that the multiple intersections of the zero-lines in Figs.~\ref{fig:irradiance_and_stokes}i-j are caused by the higher-order topological charge ($\ell=2$) of the second beam. In contrast, in Fig.~\ref{fig:irradiance_and_stokes}n, the frequencies have been swapped ($m_1=2$ and $m_2=1$), and now the circular polarization path is traced twice during the orbit's fundamental period $2\pi/\omega_0$. Thus, the double crossing in Fig.~\ref{fig:irradiance_and_stokes}o-p is due to the frequency $m_1$ and not to the topological charge. This is a fundamental difference compared to the monochromatic case where multiple intersections of the zero-lines of $S_1$ and $S_2$ can originate only from higher-order topological charges, if one beam is kept as LG$_{00}$. Such double intersection also means that each possible orientation of the orbits appears twice when tracing a closed loop around the origin. 

Further, the third bichromatic Stokes parameter $T_3$ is shown in column five, which gives the handedness of the orbits, i.e., the sense of orbit rotation. Regions where $T_3$ is positive (negative) correspond to orbits in which the electric field rotates counterclockwise (clockwise), according to the sign of the cross product of $\mathbf a$ and $\mathbf b$. The semi-minor axis $\mathbf b$ and, therefore, handedness is undefined along the $T_3$ zero-lines. The dashed lines correspond to vanishing handedness according to the alternative definition given by Eq.~(\ref{eq:h_time_circular}). Note that using $h_\mathrm{geo}$ or $h_\mathrm{ave}$ only changes the location of zero handedness, $T_3=0$ compared to $T_3=\epsilon$. 
%\EO{Note: Here, we are missing to explain anything about what those $T$ parameters tell us about the second column and how it relates to the first; overall point out similarities to what we usually see for full poincare. Plus highlight what this says about orientation and handendness (what do zerolines mean?)- why would we otherwise make the effort of deriving it at all in the first section? then you can talk about skryme. as an example: double vortex in $2\omega_0$ means each orientation twice ($T_{1,2}$), whereas single vortex in $\omega$ also gives each orientation twice which is kind of unexpected when comparing to what we know from monochromatic light ("higher order looking beam" created). maybe some of this comes later, however, I think the natural point to discuss this or at least indicate these things first, is here}

The Skyrme density $\Sigma$ is shown in the last column, where the Skyrme number $N_\text{S}$, calculated numerically over the shown area, is also depicted at the bottom. In the cases at hand, the Skyrme density has either constant sign everywhere (Fig.~\ref{fig:irradiance_and_stokes}r) or it presents two regions with opposite sign (Figs.~\ref{fig:irradiance_and_stokes}f, l; note faint blue ring in both). Qualitatively, this can be understood by looking at $T_3$. In all three rows, $T_3$ equals $+1$ at the origin, representing the north pole of the Kummer shape. When moving radially away from the origin, $T_3$ initially decreases, reaching a minimum that is $T_3\neq-1$ for the cases in the first two rows (Figs.~\ref{fig:irradiance_and_stokes}e, k) before returning to $T_3=+1$. This means that we are covering sections of the Kummer shape twice but in opposite directions, yielding $N_\mathrm S\approx 0$. On the other hand, for the case in the third row (Fig.~\ref{fig:irradiance_and_stokes}q), $T_3$ continues to decrease until it reaches $T_3=-1$ (the south pole of the Kummer shape), implying that the Kummer shape is always covered in the same direction, resulting in full coverage and a nonzero Skyrme number (here, $N_S=2$). Hence, the third row shows an example of a full Kummer beam with a skyrmionic texture.

\begin{figure}
    \centering
    \includegraphics[width=\columnwidth]{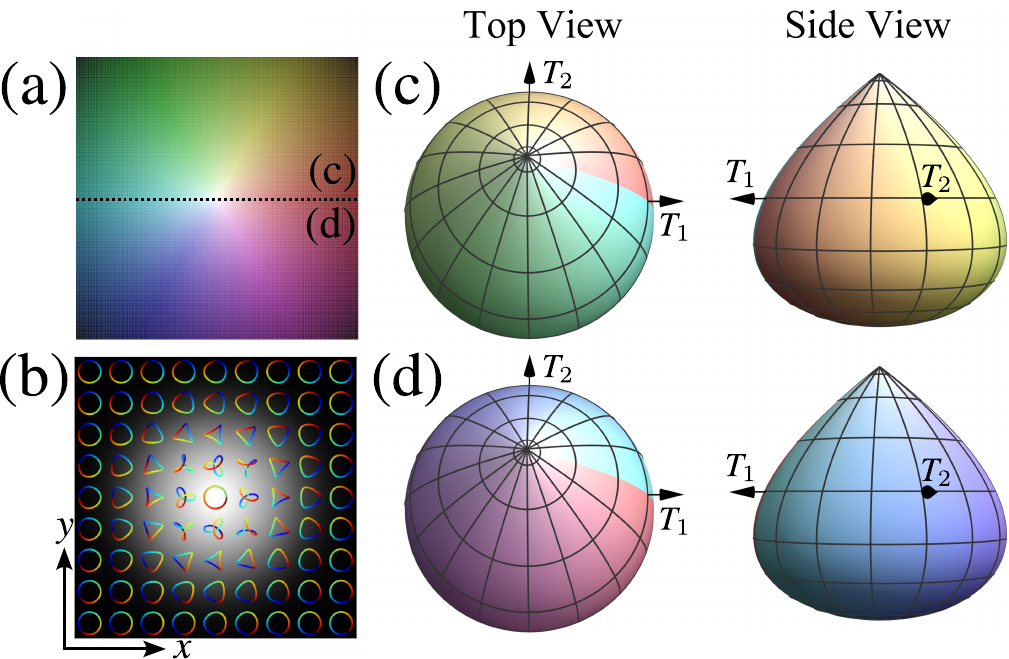}
    \caption{Coverage of Kummer shape for $m_1=2$, $m_2=1$, $\ell=1$, and equal Rayleigh ranges. (a) Colormap used to represent each point in the (b) transverse plane. Black in (a) corresponds to corners of (b), not the south pole of Kummer shapes in (c) or (d). Two Kummer shapes are necessary, corresponding to (c) the upper half-plane and (d) the lower half-plane.}
    \label{fig:projection}
\end{figure}

Let us now focus on this third example with $N_\mathrm S=2$ (third row of Fig.~\ref{fig:irradiance_and_stokes}), to gain deeper insights into how the transverse plane projects onto the Kummer shape when there is full coverage, namely, a full Kummer beam. %\JDQ{(or vice versa? Both are true obviously)}. 
When the Skyrme number is a nonzero integer, the Kummer shape is wrapped $|N_\mathrm S|$ times. Thus, this can be visualized as a mapping between the transverse plane and $|N_\mathrm S|$ copies of the Kummer shape, as shown in Fig.~\ref{fig:projection} ($m_1=2$, $m_2=1$, and $\ell=1$). An azimuthally varying color (encoding angle $\phi$) with radially varying brightness and saturation (related to radius $\rho$) is assigned to each point of the transverse plane, as shown in Fig.~\ref{fig:projection}a-b. The Kummer shapes for the upper and lower half-planes (divided by the dotted line in Fig.~\ref{fig:projection}a) are depicted in Figs.~\ref{fig:projection}c, d, respectively, from two different perspectives. In this particular case, double coverage of the Kummer shape means that its angle $\varphi$ is twice the angle $\phi$ of the transverse plane, i.e., an entire Kummer shape is spanned by only half a rotation of $\phi$. Note that this distribution matches the double crossing of the zero-lines of $T_1$ and $T_2$. Moreover, both Kummer shapes smoothly transition from white at the north pole, to color around the equator, to black at the bottom: $T_3$ is $+1$ at the origin of the transverse plane and monotonically decreases to $-1$ at infinity, matching the radial variation from red to blue in $T_3$ in Figs.~\ref{fig:irradiance_and_stokes}o-q. Note that, for visualization purposes, true ``black'' (zero brightness) was assigned to the corners of Fig.~\ref{fig:projection}(b) and not to the south pole of the Kummer shape because that requires going to infinity. %\MAA{\it maybe not needed?:}\st{How the plane is divided is arbitrary; we could have chosen the left and right half-planes instead.}

\subsection{Robustness Upon Perturbation}
%\JDQ{(What is the defect associated with the Skyrme number?)}\\
%\JDQ{(What is the relevance of the starting point of each orbit?)}
%\highlight{After reading \cite{Peters2026TopologicalLink}, I think we must calculate the Skyrme number after propagation or justify very well why it is not/cannot be calculated... Robustness of skyrmionic textures comes from the Skyrme number so bringing this up and not measure it is a bit pointless.}\\
 %(each orientation appears twice)

%Further, as pointed out above, the $T_{1,2}$ parameters indicate a higher-order-like spatial structure for the second and third row of Fig.~\ref{fig:irradiance_and_stokes} (each orientation appears twice). For monochromatic light, such higher-order fields (e.g. higher-order Poincar\'{e} beams) are known to show splitting in their zero-line intersections (i.e. monochromatic singularities) upon minor perturbation, causing changes in the spatial beam profile close to the intersections.%\MAA{\st{Exemplarily, we choose a turbulence of strength SR = 0.3 (Kolmogorov’s theory, SR: Strehl ratio \cite{Ndagano2017CharacterizingLight}).}} %which correspond to the initial
%are just the propagated fields in the last two rows of Fig.~\ref{fig:irradiance_and_stokes}. %The most significant result corresponds to %Notice that, in the second and third rows of Fig.~\ref{fig:irradiance_and_stokes} the zero-lines of $T_1$ and $T_2$ exhibit a double intersection.
Research on monochromatic structured vectorial beams has often been motivated by the robustness of the spatial distribution of the Stokes vector, which exhibits singularities that may be stable during propagation depending on the underlying phase vortices. In particular, orientation singularities are identified as intersections of the monochromatic Stokes parameters $s_1$ and $s_2$. In higher-order fields, that is, fields with points where the zero-lines cross more than once (e.g. higher-order Poincar\'{e} beams), these points are known to show splitting upon minor perturbation, causing changes in the spatial beam profile close to the intersections. Such higher-order-like spatial structure is also present in $T_1$ and $T_2$ for the second and third rows of Fig.~\ref{fig:irradiance_and_stokes}.

\begin{figure*}[t]
    \centering
    \includegraphics[width=\linewidth]{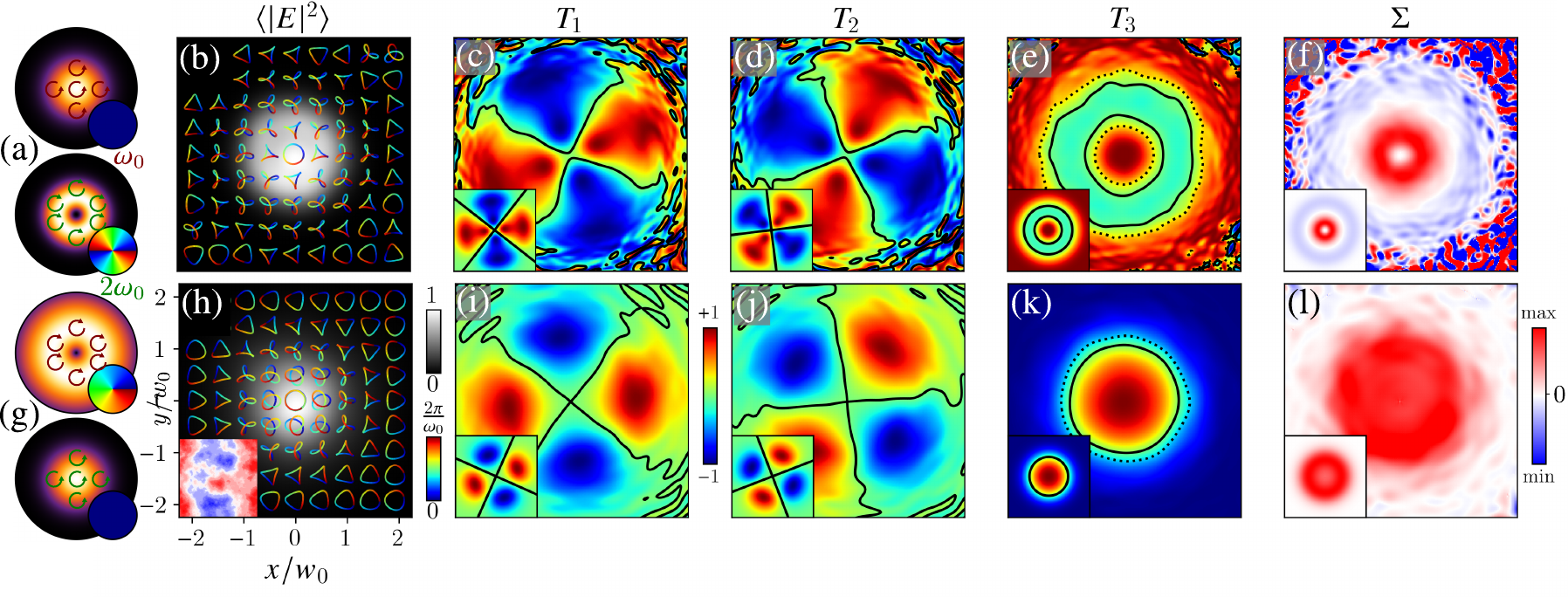}
    \caption{Effect of atmospheric turbulence during propagation for (a-d) $m_1=1$, $m_2=2$, $\ell=1$, and (e-h) $m_1=2$, $m_2=1$, $\ell=1$. Quantities across columns are the same as in Fig.~\ref{fig:irradiance_and_stokes}. Insets show the (c-f, i-l) respective unperturbed quantities and (h) applied phase turbulence.  Rayleigh ranges are equal in both cases, and the propagation distance is $z=0.25z_{\mathrm R}$. The maximum phase shift introduced by the perturbation is $\Delta\phi_\mathrm{max}\approx 0.2\pi$.}
    \label{fig:perturbation}
\end{figure*}

To understand any similarities to monochromatic structured and skyrmionic fields, we study the robustness of the topological features of bichromatic fields described above under perturbations such as those resulting from propagation through atmospheric turbulence. As exemplary results, we show the fields in the last two rows of Fig.~\ref{fig:irradiance_and_stokes} propagated through a turbulence of strength SR = 0.3 (Kolmogorov’s theory, SR: Strehl ratio\,\cite{Ndagano2017CharacterizingLight}) in Fig.~\ref{fig:perturbation} (subfigure arrangement as in Fig.~\ref{fig:irradiance_and_stokes}). The bottom left inset in Fig.~\ref{fig:perturbation}h depicts the applied phase turbulence. For reference, the insets in $T_{1,2,3}$ and $\Sigma$ show the corresponding plots at the same propagation distance but without turbulence. As discussed previously, for insets in Fig.~\ref{fig:perturbation}c-d, the double intersection can be explained by the higher-order topological charge of the second Laguerre-Gaussian beam ($\ell=2$). For the second row (see insets in Fig.~\ref{fig:perturbation}i-j), the origin is the Gaussian beam having the higher frequency ($m_1=2$). We can see that turbulence triggers the splitting of the double intersection caused by the higher-order topological charge (Figs.~\ref{fig:perturbation}c-d); this comes naturally from the underlying higher-order phase singularity, which splits upon perturbation. However, the double crossing caused by the higher-order frequency (Figs.~\ref{fig:perturbation}i-j) remains, indicating that bichromatic light requires higher-order-like topological structures that do not show the classical splitting. Notice that in both cases the intersections of $T_1$ and $T_2$ correspond to points with undefined orientation, where the associated orbits are circular; however, those in the second row of Fig.~\ref{fig:perturbation}h are a double circle, as described earlier. We can observe that the spatial distribution of the Skyrme density is fairly similar compared to the case without perturbation (cf. Figs.~\ref{fig:perturbation}f,l and their insets), except in regions near the edge of the shown area. In those regions, the irradiance is negligible and the Skyrme density exhibits very rapidly irregular oscillations (cf. Figs.~\ref{fig:perturbation}b,h and Figs.~\ref{fig:perturbation}f,l). For this reason, we do not calculate the Skyrme number in the presence of turbulence.

%\highlight{Maybe say that you can calculate the Skyrme number as a line integral...?}

%\EO{which one splits, the double or single circle? we should discuss this. Also we should say what this observation means! 1. conclusion: splitting in first row makes sense, as underlying higher order phase singularity splits upon perturbation. 2. conclusion: bichromatic light seems to allow for higher-order-like topological structures which do not show the classical splitting (however, do not talk about singularities), which will have to be investigated further in the future. (btw all of this could also come after the statement about the skyrm density, however, if miguel is sure about the "I dont think its a numerical issue" thing, we might want to add one or two sentences about skyrmions scrambling due to non-infinite energy of spatially constraint beams}

%this case because %angular spectrum propagation introduces artifacts where the irradiance is negligible but where the Skyrme density varies greatly. This could be avoided by increasing resolution and floating-point precision. \JDQ{(JDQ: I didn't understand if the following sentence is a replacement of the sentence before or an additional sentence): }the turbulent propagation introduces 
 %\MAA{(I do not think this is a numerical issue but a true scrambling of the skyrme density.)}
%%%EO: I will stop here for tonight and continue the next few days

\section{General Properties of Kummer Beams}
\label{sec:general_properties}
To provide a more quantitative understanding of the results in the previous section, we now explore how the parameters of the underlying Laguerre-Gaussian beams, such as topological charge, frequency, beam waist, Rayleigh range, and propagation distance affect the properties of Kummer beams, specifically the orientation of their orbits and their Skyrme number. In the remainder of this work, we consider the more general case of two Laguerre-Gaussian beams, forming a bicircular state, with topological charges $\ell_1$ and $\ell_2$ and radial index $n_{1,2}=0$, i.e., $E_{1}^0=\mathrm{LG}_{0\ell_1}$ and $E_{2}^0=\mathrm{LG}_{0\ell_2}$.% \JDQ{()}% By keeping the radial structure of the beams minimal, i.e., the radial index $n=0$, we are able to easily find analytical results in a simple form. %We find analytical expressions for the orientation $\psi$ and 

\subsection{Spatial Distributions of Orientation and Handedness}
\label{sec:mapping}
%\JDQ{(What happens in the more general case where planes of beam waist do not match?)}\\
Understanding how the transverse plane generally maps onto the Kummer shape during propagation requires calculating the coordinates $\varphi$ and $T_3$ of the Kummer shape [see Eq.~(\ref{eq:kummer_params})]. The angle $\varphi$, linearly related to the orientation by Eq.~(\ref{eq:orientation_circular}), is given by 
\begin{align}
\varphi(\rho,\phi,z)
    &= (m_1\ell_2-m_2\ell_1)\phi \nonumber \\
    &+m_2(|\ell_1|+1)\Phi_{\mathrm G1}(z)-m_1(|\ell_2|+1)\Phi_{\mathrm G2}(z) \nonumber \\ 
    %&+ \frac{m_2k_1(z_{\mathrm R1}^2-z_{\mathrm R2}^2)z}{2(z^2+z_{\mathrm R1}^2)(z^2+z_{\mathrm R2}^2)}\rho^2,
    &+\frac{m_1m_2\omega_0\left[R_1(z)-R_2(z)\right]}{cR_1(z)R_2(z)}\rho^2,
    \label{eq:spirals_amplitude}
\end{align}
%
%\JDQ{(mention fractional phase vortex in orientation somewhere?)}\\
%\MAA{\it Clarify:}
where we can observe, in order, terms associated with the phase vortices, Gouy phases, and wavefront curvature of each beam. The first term forms a phase vortex of order $m_1\ell_2-m_2\ell_1$ in the complex field $T_1+\mathrm i T_2$, which explains the number of intersections observed in the third and fourth columns of Fig.~\ref{fig:irradiance_and_stokes}. The last term, hence, the wavefront curvatures cause the contour lines of $\varphi$ to become spirals, with $\rho^2$ in this term quantifying how quickly these spirals twist; this effect is maximized at a propagation distance $z= \pm\left[(z_{\mathrm R1}^4+z_{\mathrm R2}^4+14z_{\mathrm R1}^2z_{\mathrm R2}^2)^{1/2}-z_{\mathrm R1}^2-z_{\mathrm R2}^2\right]^{1/2}/\sqrt{6}$. We can also see from Eq. (\ref{eq:spirals_amplitude}) that the spiral behavior disappears only at $z=0$ and $z=\pm\infty$, unless the Rayleigh ranges of the two beams are equal, which results in equal wavefront curvatures. When the Rayleigh ranges match (as was the case in Figs.~\ref{fig:irradiance_and_stokes}, \ref{fig:projection}, and \ref{fig:perturbation}), spirals disappear for all $z$, and the effect of propagation is that $\mathbf T$ experiences a rigid rotation around the $T_3$ axis. This rotation can be nulled for any pair of Laguerre-Gaussian modes if their frequencies are chosen to satisfy $\omega_2/\omega_1=(|\ell_2|+1)/(|\ell_1|+1)$. Note that Eq.~(\ref{eq:spirals_amplitude}) would still be valid for Laguerre-Gaussian beams with nonzero radial indices by performing the substitution $|\ell_{1,2}|\rightarrow|\ell_{1,2}|+2n_{1,2}$ in the second term; the radial index only changes the factor of the respective Gouy phase.

%The orientation of the polarization orbits is linearly related to $\varphi$, according to Eq.~(\ref{eq:orientation_circular}), so it has the general form
%\begin{equation}
    %\psi=\frac{m_1\ell_2-m_2\ell_1}{m_1+m_2}\phi + \psi_0(\rho,z).
    \label{eq:orientation_vs_azimuthal}
%\end{equation}
%
%\JDQ{Where is the vortex of \cite{Pisanty2019KnottingLight} exactly?} 

The handedness of the bichromatic orbits is directly related to the sign of the coordinate $T_3$ by Eqs.~(\ref{eq:h_geo_circular}) and (\ref{eq:h_time_circular}), which depends on the amplitudes of each (rotationally symmetric) beam. As we shall see in the next section, the positions where $T_3$ reaches the values $\pm 1$, i.e., the poles of the Kummer shape, are relevant when calculating the Skyrme number. In general, the north (south) pole of the Kummer shape is reached when $|E_2|/|E_1|=0$ ($|E_1|/|E_2|=0$). For Laguerre-Gaussian beams with zero radial indices, this can only occur at the origin ($\rho=0$) or at infinity ($\rho\rightarrow\infty$).

\subsection{Coverage of the Kummer Shape}
\label{sec:coverage}
%\highlight{If one makes the question: when does the light field have a Skyrminic texture? Then one would have to consider not only the entire transverse plane but also finite regions where the Skyrme number could be a (nonzero) integer, right? This problem is very complicated for $n_1,n_2>0$ but simple enough when $n_1=n_2=0$. The results below are for the general case (arbitrary $n_1$ and $n_2$) but integrating across $\mathbb R^2$, which does not reply the questions above. It might be better then to stick to the case $n_1=n_2=0$ if not all possible regions are being studied. In that case, the whole discussion around whether the boundary of the volume is included or not simplifies; it's always included. Thoughts?}\\
As observed in Sec. \ref{sec:kummer_beams}, only certain superpositions of Laguerre-Gaussian beams yield a nonzero integer Skyrme number at a given $z$ plane. This nonzero integer value indicates a nontrivial topological texture that makes the beam robust provided $N_\mathrm S$ remains constant during propagation. For a deeper understanding of this behavior, we investigate how the Skyrme number, defined in Eq.~(\ref{eq:skyrme_number}), depends on the propagation distance and the beam‑waist parameters, and whether it is an integer, zero, or otherwise. One can prove that, within an annular region limited by two rings of radii $\rho_1$ and $\rho_2$, the Skyrme number $N_\text{S}$ is simply proportional to the difference of the values of $T_3$ at the two corresponding rings, 
\begin{align}
N_\mathrm S (z)&= -\frac{1}{2}\frac{\partial \varphi}{\partial \phi} \int_{\rho_1}^{\rho_2}\mathrm d\rho\,\frac{\partial T_3}{\partial \rho} \nonumber \\
%&=\frac{m_2\ell_1-m_1\ell_2}{2}\int_{\rho_1}^{\rho_2}\mathrm d\rho\,\frac{\partial T_3}{\partial \rho} \nonumber\\
&=\tfrac{1}{2}(m_1\ell_2-m_2\ell_1)[ T_3(\rho_1,z)-T_3(\rho_2,z)].
\label{eq:skyrme_number_intermediate}
\end{align}
%\JDQ{Notice that only the derivative of $\varphi$ with respect to $\phi$ shows up and not with respect to $\rho$. This is due to the rotation symmetry of $T_3$...} %Physically, this means that the irradiance of each beam vanishes at one of the rings, regardless of which beam corresponds to which ring.  %\JDQ{[I'm pretty sure that a region that yields a (nonzero) integer Skyrme number always exists for $n_1,n_2>0$ (except if there are common zeros; math problem that I don't want to solve). Still thinking about the case where only one is zero and the other is not...]}. %\JDQ{(physical explanation to this? Notice that $\varphi$ is still a function of $\rho$ and $z$ when $z_{\mathrm R1}\neq z_{\mathrm R2}$).} 
From this equation, we can draw multiple conclusions about $N_\mathrm S$. The Skyrme number is proportional to $\partial \varphi/\partial\phi$; the dependence of the angle $\varphi$ on $\phi$, however, vanishes (e.g., only a meridian of the Kummer shape is mapped at fixed $\rho$) when $\ell_2/\ell_1=m_1/m_2$, thus making $N_\mathrm S$ identically zero. When $\ell_2/\ell_1\neq m_1/m_2$, a nonzero integer Skyrme number, implying that a skyrmionic texture exists, is only possible when $T_3(\rho_1,z)=\pm 1$ and $T_3(\rho_2,z)=\mp1$. This simply means that opposite poles of the Kummer shape are reached at radial positions $\rho_1$ and $\rho_2$ after a propagation distance $z$.
Note that this may occur in a finite region when $n_1,n_2>0$, which introduces zeros at specific radii for each monochromatic component. Finding such regions requires one to determine the $n_1$ and $n_2$ zeros of each beam, respectively, which would be functions of the beam waists $w_{1,2}$ and the propagation distance $z$. This task becomes increasingly difficult for larger radial indices and, in fact, analytically impossible for $n_1,n_2\geq 5$.
Therefore, we stay with our choice $n_1=n_2=0$. Under this simplification, the entire transverse plane is the only region of integration where $N_\mathrm S$ can be an integer. 

\begin{figure}[h]
    \centering
    \includegraphics[width=\columnwidth]{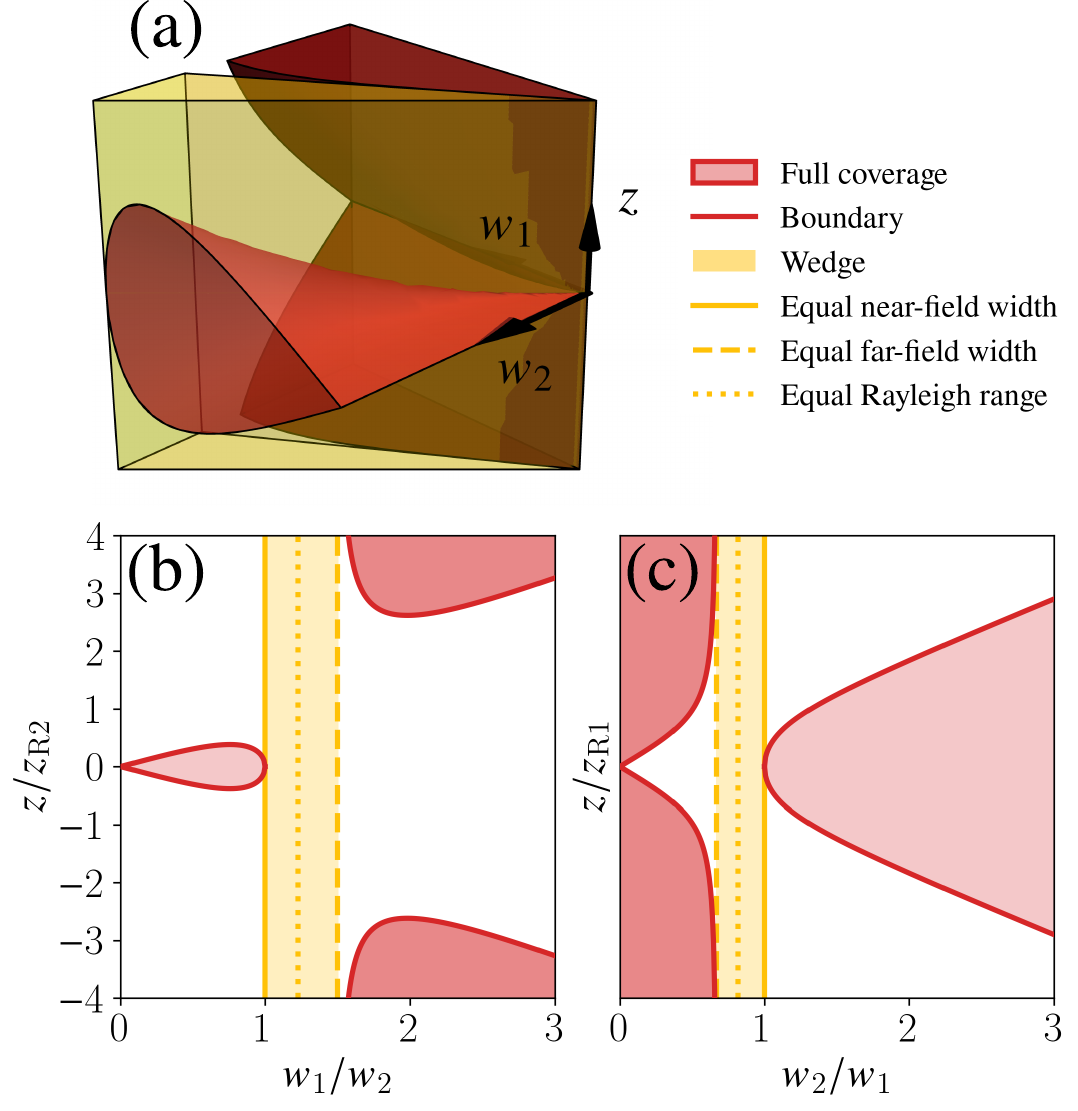}
    \caption{Full coverage region with nontrivial skyrmionic texture. (a) Representative full coverage volume $\mathcal V$ in red and wedge $\mathcal W$ in yellow for $m_1=2$, $m_2=3$, $\ell_1=0$, and $\ell_2=1$. Peak (valley) is represented by a light (dark) red. The Skyrme number equals two inside the volume and zero outside. Slices of full coverage region (red) and wedge (yellow) in terms of propagation distance and beam waist ratio when (b) $w_1$ or (c) $w_2$ is varied, while the other beam waist is kept fixed. The solid, dotted, and dashed lines correspond, respectively, to equal near-field widths, far-field widths, and Rayleigh ranges. Supplementary Visualization 2 shows the volume in (a) rotating, while Visualization 3 is an animation of the volume being sliced along the three axes $w_1$, $w_2$, and $z$}.
    \label{fig:coverage}
\end{figure}

When $|\ell_2|>|\ell_1|$, the value at the origin is $T_3(0,z)=1$. Thus, the Skyrme number is a nonzero integer only if the intensity of the $\text{LG}_{0\ell}$ beam is larger than that of the fundamental $\text{LG}_{00}$ beam. This translates to the former's beam width being equal or larger than the latter's. Quantitatively, 
$T_3(\infty,z)=-1$ therefore occurs when

\begin{gather}
    \label{eq:integer_skyrme_volume}
    \xi_2^2(z)w_2^2 \geq \xi_1^2(z)w_1^2 \nonumber\\
    \Leftrightarrow z^2(\omega_1^2w_1^2-\omega_2^2w_2^2) \geq \left(\frac{\omega_1\omega_2w_1w_2}{2c}\right)^2(w_1^2-w_2^2).
\end{gather}
%
%\MAA{where $c=\omega_i/k_i$ is the speed of light.} 
%The boundary of $\mathcal V$ may or may not have the same value of $N_\mathrm S$ as the interior depending on the relation between $n_1$ and $n_2$ because the convergence value of $T_3$ at infinity depends on the radial indices on the boundary \MAA{(\it clarify? are you saying that $N_{\rm S}$ has different values in different regions of the volume?)}.
 %We refer to $\mathcal V$, and its boundary in the appropriate case, as the full coverage region. 
 % since the inequality in Eq. (\ref{eq:integer_skyrme_volume}) does not depend on $z$  within $\mathcal W$ \MAA{\it (same question as above)}. The Skyrme number is a nonzero integer if $\omega_1>\omega_2$ and vanishes otherwise. 
 % \JDQ{(Send this paragraph to Supplementary Materials?)}. The Skyrme number vanishes if the two equalities are simultaneously satisfied, which only happens when angular indices and topological charges are respectively identical. \JDQ{(Trivial?)}
This inequality depends on the beam waists $w_{1,2}$ and the propagation distance $z$, suggesting that $(w_1,w_2,z)$ is the appropriate space to study the Skyrme number. The region defined by Eq.~(\ref{eq:integer_skyrme_volume}), which we refer to as the full coverage region, consists of a volume $\mathcal V$ (strict inequality) and its boundary $\partial\mathcal V$ (equality), where $N_\mathrm S=(m_1\ell_2-m_2\ell_1)$. This region is shown in Fig.~\ref{fig:coverage} in red for $m_1=2,m_2=3$ and $\ell_1=0,\ell_2=1$. Note that the complementary volume (everything that is not colored red in Fig.~\ref{fig:coverage}a) has a vanishing Skyrme number, implying that $N_\mathrm S$ is piecewise. This transition is depicted in Visualization 1 (Supplementary Material), where the vector $\mathbf T$ at large radii (near the border of the disk) can be seen to flip during propagation. Another region of interest in the $(w_1,w_2,z)$ space is the volume $\mathcal W$ defined by the wedge between the planes $w_2/w_1=1$ and $w_2/w_1=\omega_1/\omega_2$ (see yellow volume in Fig.~\ref{fig:coverage}a). These two planes correspond to equal beam widths for the two monochromatic components in the near field and the far field, respectively. Inside this wedge, $N_\mathrm S$ is independent of the propagation distance $z$; it is a nonzero integer if $\omega_1>\omega_2$ and vanishes otherwise ($N_\mathrm S=0$ for the case shown in Fig.~\ref{fig:coverage}a). Further, the plane $w_2/w_1=\sqrt{\omega_1/\omega_2}$, corresponding to equal Rayleigh ranges, is always inside $\mathcal W$ (in the monochromatic case $\mathcal W$ collapses onto a plane). Notice that the inequality in Eq.~(\ref{eq:integer_skyrme_volume}) and the sign of $N_\mathrm S$ would reverse if $|\ell_1|>|\ell_2|$. If $|\ell_1|=|\ell_2|$, then $N_\mathrm S$ is neither piecewise in $z$ nor an integer.
%The volumes $\mathcal V$ and $\mathcal W$ are shown in Fig.~\ref{fig:coverage}a in red and yellow, respectively, for $m_2=3>m_1=2$ and $\ell_2=1>\ell_1=0$. %\JDQ{(maybe deleting the following):\st{The peak is in contact with the wedge at $w_1=w_2$ and $z=0$, whereas the valley asymptotically reaches the wedge at $m_1w_1=m_2w_2$ and $z=\pm\infty$. The peak and valley are only connected by the origin, and the distinction between them exists only when $\omega_1<\omega_2$ since there is only one volume that includes the wedge in the opposite case $\omega_1>\omega_2$.}}
%Although in this particular case the peak and valley are clearly related, respectively, to the near and far fields, this correspondence may change since the full coverage region would be interchanged with the complement of $\mathcal V$ if $\ell_2<\ell_1$. Therefore, the roles of the peak and valley regions would be swapped and geometrically connected through $\mathcal W$, which would then be part of the full coverage region.

The full coverage region has two distinct parts, which we label for convenience as the peak (above the plane $w_1=w_2$, in light red) and the valley (below the plane $m_1w_1=m_2w_2$, in dark red), where $N_\mathrm S=2$ only at short and long propagation distances, respectively. Visualizing $\mathcal V$ is difficult, so we have included slices of the volume generated by fixing one of the beam waists. The full coverage region and wedge slices are shown in Fig.~\ref{fig:coverage}b-c in red and yellow. Each slice has two shapes, corresponding to the peak (light red) and valley (dark red), bounded by the edges of the wedge. The exact appearance of these cuts depends on the value of the fixed beam waist. As the fixed waist increases, the shapes on the left (small beam waist ratio) expand and the ones on the right (large beam waist ratio) contract. For values that are small or large enough, one of the two shapes may completely disappear. For additional illustration, we refer to Visualization 2 and Visualization 3 in the Supplementary Materials.

%The boundary of these slices is given by
%
%\begin{equation}
%    z/ z_{Ry} = \pm\left(\frac{w_x}{w_y}\right)\sqrt{\frac{(w_x/w_y)^2-1}{(w_x/w_y)^2-m_y^2/m_x^2}}
%\end{equation}
%
%where $x$ denotes the beam whose waist is varied and $y$ the one kept fixed. 

%This two-fold coverage has a different origin in each case, however. In the second row, it is caused by the high-order topological charge $(\ell=2)$, while in the third row, the reason is that the Gaussian beam has the the higher frequency $(m_1=2)$. \JDQ{should I move down this discussion???}

\section{Concluding Remarks}
Up to this point, the polarization ellipse of monochromatic beams has been one of the main tools to produce optical skyrmionic textures.
In this work, we have introduced a distinct, previously unexplored parameter space, the Kummer shape, which is completely covered by the hereby introduced full Kummer beams. Their nontrivial textures have a topological invariant governed jointly by orbital angular momentum and frequency, made explicit in $N_\mathrm S=\mathrm{sign}(|\ell_2|-|\ell_1|)(m_1\ell_2-m_2\ell_1)$. Therefore, in addition to topological charge, frequency becomes a knob to control topology, which offers a new degree of freedom with no counterpart in monochromatic light.
%For instance, a simple combination of a Gaussian an a first-order vortex already yields a higher-order Skyrme number.

Bichromaticity also changes how these textures behave under propagation and perturbation. In full Poincaré beams, a propagation-invariant Skyrme number requires equal beam waists, and any mismatch results in an abrupt change. Full Kummer beams accept beam waists inside the wedge bounded by the equal near-field width and equal far-field width planes, thus offering resilience to beam waist matching. In fact, when the beam with the lower frequency has the larger topological charge (in absolute value), the texture is nontrivial and propagation invariant. Similarly, multiple intersections in the zero-lines of $T_{1,2}$ do not split upon perturbation, proving robustness for any frequencies as long as the underlying phase vortices are first order. Therefore, bichromatic light supports structurally stable, higher-order-like orientation singularities, which are not attainable in Poincaré beams.

For bicircular fields, these features admit a direct geometric reading: $T_2/T_1$ determines the orientation of the orbit and $T_3$ quantifies its handedness, both quantities being directly related to the coordinates of the Kummer shape and defined through the generalized semi-major and semi-minor axes. Note that Freund\,\cite{Kessler2003LissajousSingularities,Freund2003BichromaticFields} had introduced alternative definitions for the orientation and handedness of Lissajous shapes in superpositions of beams with spatially varying monochromatic polarizations. However, this definition of handedness fails when the polarizations $\mathbf{\hat e}_{\mathrm R,\mathrm L}$ are spatially uniform, which is one of our assumptions, and the definition of handedness is only valid for the case $m_1=1,m_2=2$ or vice versa. Our definitions are valid for constant $\mathbf{\hat e}_{\mathrm R, \mathrm L}$ and general $m_{1,2}$.

%There are multiple points outside the scope of this paper that can be investigated in the future. For instance, the singular behavior of the higher-order-like distributions in $T_{1,2}$ requires further study. In addition, $T_3$ can be related to spin angular momentum, thus giving a physical meaning to its equipotential lines. Moreover, combining beams with orthogonal linear polarizations leads to Kummer beams possessing a different family of orbits, whose geometrical and physical properties are related differently to the bichromatic Stokes parameters. Finally, this theory can be generalized to partially polarized light as well as to non-paraxial light.

Beyond their theoretical interest, these features open practical directions. Bicircular $\omega$--$2\omega$ fields, the case $m_1=1$, $m_2=2$ (or vice versa) studied here, are used in strong-field and attosecond physics to generate circularly polarized high harmonics and to tailor chiral light--matter interactions; full Kummer beams would provide control over the local orientation and handedness of the field \cite{Smirnova2025ASensing,Tyulnev2024ValleytronicsField,Heinrich2021ChiralFields}. In the context of communication, where skyrmionic textures have been suggested to be used for topological information encoding\,\cite{Bezuidenhout2026DeepTraining}, frequency can act as an additional degree of freedom of the mapped domain\,\cite{Vo2024ClosedPulses,Vo2026AgileTextures,Ornelas2026TunableMultiplexing,Teng2025ConstructionSkyrmions}. Here, frequency directly influences the target space and enables multiplexing, thus increasing efficiency and channel capacity. Assessing this potential will require quantifying $N_\mathrm{S}$ under realistic perturbations, which lies beyond the scope of this work. Further open questions include the singular structure of the higher-order-like distributions in $T_{1,2}$, the use of orthogonal linear polarizations, whose orbits relate differently to the bichromatic Stokes parameters, and the extension to partially coherent and non-paraxial light.

\begin{backmatter}
    
\bmsection{Funding}
This research received funding from the
Agence Nationale de la Recherche (ANR) through the project 3DPol, ANR-21-CE24-0014-01.

\bmsection{Acknowledgment}
The authors thank David Marco for useful discussions. E.O. acknowledges startup funds from the University of Rochester.

\bmsection{Disclosures}
The authors declare no conflicts of interest.

\bmsection{Data availability} No data were generated or analyzed in the presented research.

\bmsection{Supplemental document}
See Supplement 1 and Visualizations 1, 2, and 3 for supporting content.

\end{backmatter}
% Bibliography
\bibliography{references}
\bibliographyfullrefs{references}

% Full bibliography added automatically for Optics Letters submissions; the following line will simply be ignored if submitting to other journals.
% Note that this extra page will not count against page length

%Manual citation list
%\begin{thebibliography}{1}
%\bibitem{Zhang:14}
%Y.~Zhang, S.~Qiao, L.~Sun, Q.~W. Shi, W.~Huang, %L.~Li, and Z.~Yang,
 % \enquote{Photoinduced active terahertz metamaterials with nanostructured
  %vanadium dioxide film deposited by sol-gel method,} Opt. Express \textbf{22},
  %11070--11078 (2014).
%\end{thebibliography}

% Please include bios and photos of all authors for aop articles

\end{document}